\documentclass[twocolumn,showpacs,aps]{revtex4}
\usepackage{bm,color,amsmath,amssymb,mathrsfs,latexsym,graphicx,psfrag,dsfont}
\graphicspath{{./}{./pics/}}

\newcommand{\bs}[1]{\boldsymbol{#1}}
\newcommand{\vac}{\left|\,0\,\right\rangle}
\newcommand{\up}{\uparrow}
\newcommand{\dw}{\downarrow}

\newcommand{\ket}[1]{\left|#1\right\rangle}

\def\ie{i.e.,\ }
\def\eg{e.g.\ }
\def\z{\text{z}}

\begin{document}

\title{Fractional spin liquid hierarchy for spin $\bs{S}$ antiferromagnets}
\author{Burkhard Scharfenberger$^1$, Ronny Thomale$^2$ and Martin Greiter$^1$}
\affiliation{${}^1$Institut f\"ur Theorie der Kondensierten
  Materie and DFG Center for Functional Nanostructures (CFN), KIT,
  Campus S\"ud, D 76128 Karlsruhe}
\affiliation{${}^2$Department of Physics, Princeton University,
  Princeton, NJ 08544}

\date{\today}
\pagestyle{plain}

\begin{abstract}
We propose a hierarchy of spin liquids with spin $S$, which are
generated via a Schwinger boson projection from $2S$ (Abelian) chiral
spin liquids with $S=1/2$.  We argue that the P and T invariant
liquids we construct for integer spin $S$ support SU(2) level $k=S$
non-Abelian statistics.  We conjecture that these spin liquids serve
as paradigms for antiferromagnets which are disordered through mobile
charge carriers.
\end{abstract}


% 75.10.Kt  Quantum spin liquids, valence bond phases and related phenomena 
% 75.10.Jm  Quantized spin models, including quantum spin frustration 
% 05.30.Pr  Fractional statistics systems (anyons, etc.) 

\maketitle

%\section{Introduction \label{sec:intro}}

\emph{Introduction.}---Topological phases have become a major branch
of research in condensed matter
physics~\cite{wen-89prb7387}. Intriguing features of a new kind of
order which is not characterized by a broken symmetry (and hence a
local order parameter), but depends on global properties of the
system, have been investigated in a plethora of different models.  The
first fundamental understanding of a topological phase has been
accomplished in the quantum Hall effect~\cite{wen-90prb9377}. There,
the formation of quantized plateaus in the off-diagonal conductivity
is connected to the manifestation of gapless edge states in a
bulk-gapped system which are robust against local perturbations of the
two-dimensional electron gas.  Many new areas where topological phases
have been discovered are intimately linked to the quantum Hall effect.
Kane and Mele~\cite{kane-05prl226801} recently opened up the field of
topological insulators in two spatial dimensions, when they
beautifully generalized a graphene fractional quantum Hall (FQH) model
due to Haldane~\cite{haldane88prl2015} by introducing spin and a
spin-orbit coupling term which exhibits a quantum spin Hall effect
with non-chiral edge states.

The concept of fractional quasiparticle
statistics~\cite{wilczek90,stern10n187} as another possible
manifestation of topological order, has been formulated first in the
context of the quantum Hall
effect~\cite{halperin84prl1583,arovas-84prl722}.  In addition to
Abelian statistics, where the many-particle wave function acquires a
fractional phase upon braiding of quasiparticles~\cite{wilczek90},
certain systems exhibit non-Abelian fractional statistics, where the
braiding induces rotations in the topologically degenerate ground
state manifold~\cite{stern10n187}.  The paradigm for a state with
non-Abelian statistics is the Pfaffian
state~\cite{moore-91npb362,greiter-92npb567}, which has been proposed
to describe the FQH plateau at $\nu=5/2$~\cite{greiter-92npb567}.  The
quasihole excitations of this state directly relate to topologically
protected vortex core zero modes of certain charged and neutral
superfluids such as chiral superconductors and superfluid
${}^3\text{He}$~\cite{volovik99jetp601,read-00prb10267}.

% This property could be employed for topological quantum
% computation~\cite{nayak-08rmp1083}.

Spin liquids are another diverse area where topological phases
appear~\cite{lee08s1306}.  There, quantum fluctuations manage to
suppress any formation of magnetic or other kind of local (\eg spin
solid) order, which would break the continuous symmetries, such as
global SU(2) spin or the translational symmetry, of the
system~\cite{balents10n199}.  Spin liquids have attracted particular
attention in the context of high-$T_c$
superconductivity~\cite{anderson87s1196}, and are now established as a
major playground for new developments in field theory and quantum
criticality~\cite{sachdev08np173}.  Whereas previously mentioned
examples such as the quantum Hall effect or topological insulators
generally promise a rather accessible detection of topological order
through the classification of the edge states, the situation is more
subtle for spin liquids. While they do not exhibit gapless edge states
in general, spin liquids still possess hidden topological order
emergent through fractionalization of quantum numbers and
quasiparticle statistics.  In fact, fractional quantum numbers have
been first observed in polyacetylene where the elementary spin
excitations were found to carry not $S=1$, but
$S=1/2$~\cite{su-79prl1698}.  One particular, analytically well
defined approach to the field of spin liquids is motivated by the
quantum Hall effect.  Based on an idea by D. H. Lee, Kalmeyer and
Laughlin~\cite{kalmeyer-89prb11879} proposed and developed the
(Abelian) chiral spin liquid (CSL)~\cite{wen-89prb11413}, which in
essence is a bosonic quantum Hall wave function of spin flip
operators.  The elementary excitations are spinons, which carry
$S=1/2$, no charge, and obey half-Fermi statistics.  While the CSL did
not turn out to be directly relevant to high-$T_c$ superconductors, it
is the system in which Wen~\cite{wen-89prb7387} first established the
basic notion of topological order by computing its topological
degeneracy.  The low energy physics is accordingly described by a
Chern--Simons theory.  Further independent branches diversified the
field of spin liquids, such as the $Z_2$ spin liquids which, in
addition to spinons, possess vison excitations which carry no spin but
couple to an emergent $Z_2$ gauge field.  $Z_2$ liquids appear to be
good candidates for certain frustrated $S=1/2$ antiferromagnets and
dimer models~\cite{moessner-01prb024504,xu-09prb064405}, and further
describe some prominent models in which topological phenomena can be
discussed on an analytical footing.  These models include Kitaev's
toric code and honeycomb models~\cite{Kitaev06ap2}.

In this Letter, we use the CSL as a building block to generate a
hierarchy of SU(2) invariant quantum spin liquids as candidate states
for disordered spin $S$ antiferromagnets.  The reason there is no
order in these systems could be frustration, but we believe that our
constructions are more likely to be realized in antiferromagnets with
itinerant holes, as the holons one would obtain in the liquids are
likely to be vastly more mobile than in competing models.  Relating
CSLs to a quantum antiferromagnet, the ostensible drawback is that, as
opposed to the $Z_2$ spin liquids, the CSL breaks parity (P) and
time-reversal (T), while antiferromagnets generically conserve these
symmetries on a Hamiltonian level and should only exhibit spontaneous
breaking of P and T in very peculiar cases.  For integer spin $S$,
this obstacle can be circumvented by combining a total of $2S$ CSLs
with opposite chiralities via a Schwinger boson projection
technique~\cite{greiter02jltp1029}.  The resulting states hence retain
the topological features encoded in the CSL and additionally restore
the symmetries P and T, establishing a new promising route to define
spin liquid candidate states for disordered quantum
antiferromagnets. As one particularly interesting example, we propose
and investigate an $S=2$ liquid which conserves P and T and exhibits
non-Abelian spinon excitations, and so opens up a new perspective on
non-trivial topological order in higher spin antiferromagnets.

\emph{Chiral spin liquid.}---Let us define the fundamental building
block of our spin liquid hierarchy construction.  Consider a rhombic
$S=1/2$ lattice in the complex plane, $\eta_{n,m}=na+mb$, where
$n,m\in\mathds{Z}$, $a,b\in\mathds{C}$, and
$\Im b\bar a=1$, where $\Im$ denotes the imaginary part.  The CSL
state for a circular droplet with $N$ sites is given by
\begin{align} 
  \ket{\Psi_+^\mathrm{CSL}}= \sum_{\left\{z_i\right\}}
  \psi_+^\mathrm{CSL}(z_1,\ldots,z_M) S^+_{z_1}\dots S^+_{z_M}\ket{\dw}_N,
\end{align}
where we use complex coordinates $z_i$ to denote the positions of the
$M=N/2$ spin flip operators 
% and where the sum extends over all possible subsets of $M=N/2$ sites
on the lattice. $\ket{\dw}_N$ is the fully spin-down polarized state,
\ie $\ket{\dw}_N=\ket{\dw\dots\dw}$, and ${\psi}_+^\mathrm{CSL}$ is
the CSL wave function
\begin{align}\label{cslwave} 
  {\psi}_+^\mathrm{CSL}(z_1,\ldots,z_M)=\prod_{j<k}^M(z_j-z_k)^2\,
  \prod_{j=1}^M G(z_j)\,\mathrm{e}^{-\frac{\pi}{2}|z_j|^2}.
\end{align}
Here $G(\eta_{n,m})= (-1)^{(n+1)(m+1)}$ is a gauge factor which is
valued $-1$ on the doubled unit cell superlattice and $1$ otherwise
and is needed to assure the singlet property of the CSL. Up to
$G(z_j)$, the form invariance to a bosonic FQH Laughlin droplet wave
function at Landau level filling fraction $\nu=1/2$ is apparent, \ie
the CSL can be interpreted as bosons at half filling in a fictitious
magnetic field of one Dirac quantum per plaquet.  The CSL possesses
chirality ``$+$'', \ie $\langle\chi\rangle=\langle
\bs{S}_i(\bs{S}_j\times\bs{S}_k)\rangle>0$, where the sites $i,j,k$
span a triangle which is oriented such that
$\Im\{(\eta_k-\eta_i)/(\eta_j-\eta_i)\}>0$.  As $\chi\rightarrow
-\chi$ under P and T, the CSL \eqref{cslwave} breaks P and T.  A CSL
of opposite chirality ``$-$'' is obtained by complex conjugation,
$z \rightarrow \bar{z}$.  The spinon excitations are the analog of the
quasiholes in the quantized Hall states, and obey half-fermion
statistics $\theta=\pm \pi/2$, with the sign depending on the
chirality.  This also applies to holon excitations when the CSL itself
is studied away from half filling of electrons.  The CSL can be
generated alternatively through Gutzwiller projection of filled Landau
levels for $\up$ and for $\dw$ spins, and hence be defined on any
lattice
type~\cite{zou-89prb11424,laughlin-90prb664,greiter02jltp1029}.

\emph{Topological degeneracy.}---Aside from the fractional statistics
of the spinons, the non-trivial topology of the CSL manifests itself
in a degeneracy on a non-trivial manifold such as the torus, which we
obtain by imposing periodic boundary conditions.  There are various
ways to determine this degeneracy for CSLs.  Given any ground state
wave function, one way would be to obtain the topological entanglement
entropy~\cite{kitaev-06prl110404,levin-06prl110405}.  With the
computational facilities presently available, however, we found it
numerically challenging to write out the state vectors of hierarchy
states with higher spin $S$ for sufficiently large clusters.  A more
viable approach is to exploit the closed analytic
expression~\eqref{cslwave} of the CSL ground state wave function.  For
the torus, topological degeneracies (TDs) arise as we demand
quasi-periodicity under magnetic translations of a single particle
along one of the principal directions $\hat{1}$ and $\hat{\tau}$,
which span the (parallelogram-shaped) {principal region}
$\mathscr{P}_\tau=\left\{u+v\tau: u,v\in [0,1[ \subset\mathds{R},
    \Im\tau>0\right\}$.  To formulate the CSL on the
    torus~\cite{haldane-85prb2529}, we replace the factors
    $(z_i-z_j)^2$ by $\vartheta_{\mbox{-\,\!-}}^2(z_i-z_j|\tau)$,
    where $\vartheta_{\mbox{-\,\!-}}$ is the odd Jacobi theta
    function~\cite{mumford83}.  Demanding that the wave function
    is strictly invariant under translations of lattice sites by ${1}$
    and ${\tau}$, one finds that in order to cancel any position
    dependence in the phase factors, one has to introduce a set of $2$
    parameters stemming from the power of the Jastrow factor, the
    center-of-mass-zeroes $Z_\nu$. Their number, $2$, then corresponds
    to the TD of the ``toroidalized'' state and is the ground state
    degeneracy of a Hamiltonian with a suitable potential to stabilize
    the CSL on a torus
    geometry~\cite{schroeter-07prl097202}. %,thomale-09prb104406}.
    Numerically, when we write down the CSL wave functions for
    different values of the parameters $Z_\nu,\,\nu=1,2$, they will
    span a two-dimensional functional space
\begin{align}
  \mathrm{dim}\left\{\ket{\psi^{\mathrm{CSL}}_\pm\{Z_\nu\}}:
  Z_\nu\in \mathscr{P}_\tau \right\} = 2 .
\end{align}
This is the scheme we will use to determine the TD of all hierarchy
states below.

\emph{Schwinger boson projection.}---In order to construct the spin
liquid hierarchy, we first rewrite the CSL wave function 
in terms of Schwinger bosons~\cite{greiter02jltp1029,greiter-09prl207203}:
\begin{align}
  \ket{\psi_+^\mathrm{CSL}} =  \hat{\Psi}_+^\mathrm{CSL}[a^\dagger,b^\dagger]\ket{0},
\end{align}
where
\begin{align}
 \hat{\Psi}_+^\mathrm{CSL}[a^\dagger,b^\dagger] &= \sum_{\left\{z_i; w_j\right\}}
     {\psi_+^\mathrm{CSL}}(z_1,\dots,z_M) 
     \nonumber\\[-.3\baselineskip]
     &\quad\hspace{50pt} \cdot a^\dagger_{z_1}\dots
     a^\dagger_{z_M}b^\dagger_{w_1}\dots b^\dagger_{w_M},
\end{align}
and the sum extends over all possible distributions of disjoint sets
of sites $\left\{z_1,\dots,z_M\right\}$ and $\left\{
w_1,\dots,w_M\right\}$ on the lattice.  The operators $a^\dagger_i$
and $b^\dagger_j$ create auxiliary bosons, with the correspondence
between a spin $S,S^\z$ state and a state with $2S$ Schwinger bosons
is given by
\begin{align}
  \ket{S,S^z}
  =\frac{(a^\dagger)^{S+S^\z}(b^\dagger)^{S-S^\z}}{\sqrt{(S+S^\z)!(S-S^\z)!}}
  \ket{0}.
\end{align}

\begin{figure}[t]
 \centering
 \includegraphics[width=.48\textwidth]{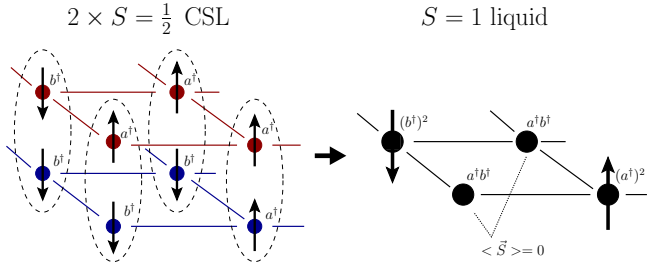}
 \caption{(color online) We create a spin $S\!=\!1$ liquid out of two
   $S\!=\!1/2$ CSLs by projection onto the symmetric spin
   configuration on each site.  Schwinger bosons ensure this is done
   automatically when multiplying the liquids. }
 \label{fig:one}
\end{figure}

\emph{$S=1$ liquids.}---Let us first study the case of constructing
$S=1$ spin liquids from projection of two CSLs. In the Schwinger boson
basis, we can accomplish the projection of the $S=1/2$ CSLs by simply
multiplying their creation operators, as illustrated in
Fig.~\ref{fig:one}.
Since the operators $a^\dagger_i, b^\dagger_j$ are bosonic, no
re-ordering is needed.  When we merge two CSLs, the resulting onsite
terms of the form $(a^\dagger_i)^2,a^\dagger_i b^\dagger_i$ or
$(b^\dagger_i)^2$ inherently belong to the symmetric $S=1$ part of the
onsite product space $\frac{1}{2}\,\otimes\,\frac{1}{2}$.  We can build
two different $S=1$ liquids out of two $S\!=\!1/2$ CSLs, depending on
whether they have equal or opposite chirality.  In the first case, we
obtain the $S=1$ non-Abelian chiral spin liquid
(NACSL)~\cite{greiter-09prl207203},
\begin{align}
 \ket{\psi^\mathrm{NACSL}_+} 
 = \left(\hat{\Psi}_+^\mathrm{CSL}[a^\dagger,b^\dagger]\right)^2\vac.
 \label{eq:s1b}
\end{align}
The NACSL still breaks P and T, but exhibits (Ising-type) non-Abelian
spinon excitations, as it resembles the bosonic Pfaffian state at
Landau-level filling $\nu=1$.  To obtain its TD, we create different
liquids parameterized by the CSL center-of-mass zeros before projection
and compute the dimension of the functional space.  As listed in
Tab.~\ref{tab:one}, the NACSL has a TD of $3$.  Since the TD is lower
than $4$, \ie lower than the product of the two constituent CSL
liquids, there is a blocking mechanism at work.  This indicates that
the statistics is non-Abelian~\cite{oshikawa-06prl060601}.
\begin{table}[t]
 \renewcommand\arraystretch{1.2}
 \begin{tabular}{p{.06\textwidth}lc}
   \emph{spin}& \emph{liquid} & \emph{top. degeneracy} \\
   \hline
   $1/2$ & $\mathrm{CSL}$                         & 2 \\
   $1$   & $(\mathrm{CSL}_+)^2$                   & 3 \\
         & $\mathrm{CSL}_+\mathrm{CSL}_-$         & 4 \\
   $3/2$ & $(\mathrm{CSL}_+)^3$                   & 4 \\
         & $(\mathrm{CSL}_+)^2\mathrm{CSL}_-$     & 6 \\
   $2$   & $(\mathrm{CSL}_+)^4$                   & 5 \\
         & $(\mathrm{CSL}_+)^2(\mathrm{CSL}_-)^2$ & 9 \\
   $k/2$ & $(\mathrm{CSL}_+)^k$                   & $k+1$ \\
 \end{tabular}
 \caption{Topological degeneracies a different hierarchy liquids. The
   liquids have been investigated numerically up to 16 sites on a
   square lattice by varying the center-of-mass-zeros to probe the
   space of degenerate states.}
 \label{tab:one}
\end{table}

Projecting two CSLs of opposite chirality gives the spin $S=1$ chirality
liquid (S1CL)~\cite{greiter02jltp1029},
\begin{align}
  \ket{\psi^\mathrm{S1CL}}
  = \hat{\Psi}_+^\mathrm{CSL}[a^\dagger,b^\dagger]
  \,\hat{\Psi}_-^\mathrm{CSL}[a^\dagger,b^\dagger]\vac.
  \label{eq:s1a}
\end{align}
The resulting wave function is real, implying the absence of a net
chirality and thus invariance under the action of P and T.  We obtain a
TD of $4$.  The absence of blocking suggests that the spinons keep
their statistical properties from the CSLs, but now carry a chirality
quantum number.  Due to the invariance under P and T, this spin liquid
is an interesting trial state for $S=1$ antiferromagnets which, as
mentioned already, can be defined on any lattice.

\emph{Spin liquid hierarchy.}---We now generalize this construction to
arbitrary spin $S$, \ie to a hierarchy of states obtained by
projecting an arbitrary number of CSLs.  Specifically, we form spin
$S=k/2,\ k\in\mathds{N}$ liquids from $k$ copies of the CSL as the
basic building blocks.  If we constrain ourselves to just one type of
chirality, the wave functions of the spin $S$ liquids,
\begin{align}
  \psi^{k-\mathrm{CSL}}_{+} 
  = \left(\hat{\Psi}_+^\mathrm{CSL}[a^\dagger,b^\dagger]\right)^k \vac,
  \label{k-CSL}
\end{align}
assume the functional form of bosonic Read-Rezayi quantum Hall states
at Landau level filling fractions $\nu=k/2$~\cite{read-99prb8084}.

Using conformal field theory, one finds that the TD of the $k$-th
Read-Rezayi states is given by $k+1$~\cite{wen-98prb15717}, which
matches the results we obtain numerically (see Tab.~\ref{tab:one}).
For $S=3/2$ or any other half-integer spin, we cannot form a P and T
invariant liquid. Taking two CSLs of one chirality and one CSL of the
other chirality gives a liquid with a $6$-fold TD.  The construction
suggests that this liquid supports spinons with (Ising-type)
non-Abelian statistics in the chiral sector where two CSLs have been
projected together and spinons with Abelian half-Fermi statistics in
the other chirality sector.

\emph{$S=2$ chirality liquid.}---A particularly interesting spin
liquid trial state can be constructed for $S=2$.  Taking two CSLs of
each chirality before projection, we obtain a P and T invariant
liquid.  We can construct the state either from two S1CLs, both of
which support Abelian spinons and are P and T invariant, or from two
\mbox{NACSLs} with (Ising-type) non-Abelian spinons and opposite
chirality,
\begin{align}
  \ket{\psi^\mathrm{S2CL}}
  &=\left(\hat{\Psi}^\mathrm{S1CL}[a^\dagger,b^\dagger]\right)^2\vac
  \nonumber\\[.5\baselineskip]
  &=\hat{\Psi}^\mathrm{NACSL}_+[a^\dagger,b^\dagger]\,
  \hat{\Psi}^\mathrm{NACSL}_-[a^\dagger,b^\dagger]\vac .
  \label{s2cl}
\end{align}
The TD of the final state is $9$-fold (see Tab.~\ref{tab:one}).  It is
hence the first instance of a blocking mechanism for a non-chiral
hierarchy state, which reduces the TD from $4\cdot 4=16$ for the
constituent S1CLs to $9$.  The construction as well as the TD suggests
that~\eqref{s2cl} exhibits (Ising-type) non-Abelian spinon statistics
for both chiralities.

The $S=2$ chirality liquid state \eqref{s2cl} is a promising candidate
to capture a universality class of disordered $S=2$ antiferromagnets,
where the spin liquids may be stabilized through itinerant holes of
appropriate kinetic energies, as the holon excitations in the
hierarchical spin liquids presumably share the very high mobility of
the holons of the individual constituent CSLs.  The characteristic features
of this universality class are, first, a $(S+1)^2$ fold TD for the P and T
invariant spin $S$ hierarchy liquid on the torus, and second, that the
spinons and holons obey non-Abelian SU(2) level $k=S$ statistics.  The
properties of the spinons would manifest itself not only in the spin
liquid state, but in the response to all probes which measure energy
scales beyond the ordering temperature, such as \eg Raman scattering.

\emph{State counting and effective field theory.}---The are important
subtleties associated with the Schwinger boson projection scheme we
employ here to obtain the hierarchy of spin liquids. The Hilbert space
of an $N$-site spin $S=1/2$ lattice contains $2^N$ states and is
matched by the number of states in the configuration space of the
spinon excitations one can create in the CSL.  For higher spin,
however, the Schwinger boson projection maps a $2^{N k}$ dimensional
product space to a $(k+1)^{N}$ dimensional space, in which the
resulting spin liquid is defined. This poses the fundamental problem
of how this reduction manifests itself for the spinons in the
hierarchy liquids. For example, while the $S=1$ chirality
liquid~\eqref{eq:s1a} suggests a picture of free spinons of different
chiralities, this picture can only be true at the lowest energies,
since the state counting does not match.  It will be interesting to
investigate this issue further from a field theoretical perspective.
Starting with an effective Chern--Simons theory for a single chiral
spin liquid~\cite{wen-89prb11413}, and the appropriate generalizations
for the higher spin CSLs \eqref{k-CSL}, we are led to conjecture that
doubled Chern--Simons theories with the appropriate $k$ may describe
the low energy physics for the non-Abelian higher spin chirality
liquids.  It is possible, however, that these theories have to be
constrained, even though the number of consistent choices appears to
be limited~\cite{hansson-04ap497}.

\emph{Conclusion.}---We propose a hierarchy of spin liquids, which is
constructed from chiral spin liquids as fundamental building blocks.
We have shown that in the P and T invariant liquids we obtain for spin
$S=2$ and higher, the topological degeneracies is reduced from what
one would expect from the product of the constituent liquids.  This
provides an indication that the statistics of the spinons is
non-Abelian.  We present arguments indicating that the statistics for
the P and T invariant spin $S$ liquid is SU(2) level $k=S$.  Finally,
we conjecture that it might be possible to stabilize these liquids
through doping with itinerant charge carriers.

% A numerical verification of this conjecture is one of the main
% projects we are presently investigating.

RT is grateful to T.H.~Hansson and S.L.~Sondhi for insightful
discussions.  BS was supported by the Landesgraduiertenf\"orderung
Baden-W\"urttemberg.  RT was supported by the Humboldt Foundation and
DFG SPP 1458/1.

\bibliographystyle{prsty}
%\bibliography{paper}
%\bibliography{../../bib/paper,../../bib/book,../../bib/htc}

\end{document}